# Atomistic Non-equilibrium Green's Function Simulations of Graphene Nano-Ribbons in the Quantum Hall Regime

Roksana Golizadeh-Mojarad*, A. N. M. Zainuddin*, Gerhard Klimeck, Supriyo Datta
*These authors contributed equally to this work*
*Network for Computational Nanotechnology, School of Electrical and Computer Engineering, Purdue University, West Lafayette, IN 47907.*
E-mail: rgolizad@purdue.edu, azainudd@purdue.edu

**Abstract.** The quantum Hall effect in Graphene nano-ribbons (GNR) is investigated with the non-equilibrium Green's function (NEGF) based quantum transport model in the ballistic regime. The nearest neighbor tight-binding model based on $p_z$ orbital constructs the device Hamiltonian. GNRs of different edge geometries (Zigzag and Armchair) are considered. The magnetic field is included in both the channels and contact through Peierls substitution. Efficient algorithms for calculating the surface Green function are used to save computation time while simulating realistically large dimensions comparable to those used in experiments. Hall resistance calculations exactly reproduce the quantum Hall plateaus observed in the experiments. Use of large dimensions in the simulation is crucial in order to capture the quantum Hall effect in magnetic fields within experimentally relevant 10–20T.

## 1. Introduction

Graphene, a two dimensional (2D) array of carbon atoms arranged in a honeycomb lattice, has recently drawn a lot of interest after being extracted through a technique called micromechanical cleavage [1, 2]. Due to its special bandstructure, this new material has not only a high mobility but also a lot of other extraordinary properties. The linear dispersion at the two inequivalent K and K′ valleys implicates a lot of quantum mechanical effects for which Graphene has attracted much experimental and theoretical interest. This opens up new possibilities in the field of microelectronics, spintronics, semiconductors and other novel applications [3].

One of the surprising properties in Graphene is its unconventional quantum Hall effect (QHE). In Graphene, the Hall resistance ($R_H$) shows plateaus at $h/(2e^2)$, $h/(6e^2)$, $h/(10e^2)$…etc for two valleys and two spins with increasing magnetic (B) field. In other words, the Hall conductivity ($\sigma_H$) is increasing with odd integer filling factor of ($e^2/h$) quantum conductance unit per valley per spin. A lot of theoretical work has been done to explain the origin of such unconventional sequence in the Hall plateaus in Graphene [4, 5]. In those works the phenomenon has been explained from a continuum model of edge states. It has been identified that there exists a zeroth Landau Level (LL) in addition to two fold valley degenerate LLs in Graphene that originates such an odd integer QHE.

This work examines the QHE with the NEGF quantum transport model for Graphene nano-ribbon (GNR) systems. Several numerical challenges in simulating a large ribbon structure to keep B-field values within the experimental range are discussed in detail. The zeroth LL contributes an additional conducting edge mode by splitting up (down) in energy for electrons (holes) near the edges in the presence of a hardwall confining potential. If the Fermi Levels ($E_F$) is positioned in between two LLs an odd number of conducting edge modes giving rise to the odd integer filling of Hall plateaus.

## 2. Theory

The Non-equilibrium Green's function (NEGF) quantum transport model [6] within the ballistic regime is used. The Hamiltonian (H) for Graphene nanoribbons (GNR) of different edges, Zigzag edge (ZGNR) and Armchair edge (AGNR) is constructed with in the atomistic nearest-neighbor $p_z$-orbital tight-binding model. Fig. 1, shows the 2D simulation structure in the NEGF formalism. Both the channel and the contact regions have the same edge geometries and width (W) in this work. The presence of the B-field applied perpendicular to the 2D plane of GNR is incorporated directly into H through the electronic coupling between nearest neighbors with the of Peierls substitution. If **A** is the vector magnetic potential where ($\nabla \times A = B$), then the electronic coupling $t_{ij}$ between sites $r_i$ and $r_j$ becomes, $t_{ij} = t_0 \exp(ie\mathbf{A} \cdot (\mathbf{r_i} - \mathbf{r_j})/\hbar)$. Here $t_0$ (~ 2.71eV) is the overlap integral between nearest carbon atoms. *e is* electron charge and $\hbar$ is Planck's constant.

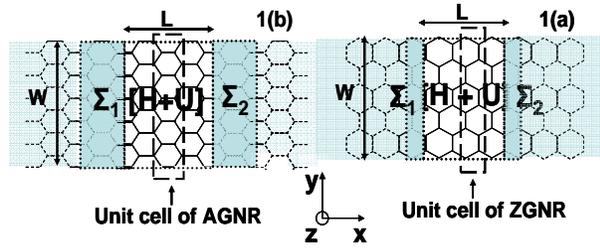

*Figure.1. Schematic illustration of NEGF quantum transport model, a) ZGNR, b) AGNR. Unit cells are shown with the dashed lines.*

Standard NEGF equations for calculating device Green's function and other relevant quantities of interest are solved according to the formalism prescribed in [7]. As Hall voltage, we calculate the quasi-Fermi level difference of the opposite edge of the channel from the ratio of electron density and spectral function. The contact self-energy matrices $[\Sigma_{1,2}(E)]$ are calculated through the surface Green function (SGF) technique [7] calculated from the following equation,

$$g_s = [(EI - \alpha_c) - \beta_c g_s \beta_c^+]^{-1} \quad (1)$$

where $g_s$ = SGF, $\alpha_c$ is the unit cell Hamiltonian and $\beta_c$ is the interlayer coupling matrix in the contacts. Practical devices are typically several micrometers long and the magnetic field extends over the whole device. The simulation of such an explicit device dimension would be computationally prohibitively expensive. One of the powers of the NEGF formalism is its ability to collapse large device regions into a single self-energy [6, 8]. In the past [9], we have developed self-energies for 2D-waveguide structures without B-field. With such a boundary condition one would have to spatially ramp up the B-field very smoothly toward the center of the device, still requiring a significant device length. Here we developed a surface self-energy with constant B-field which then allows us to simulate the central device width a single central unit cell. The boundary self-energy captures an infinite number of unit cells.

To compute the SGF we employ two numerical algorithms. The first one is the simplest iterative method where the iteration starts from a reasonable guess and recursively updates the value of SGF by going down the layers of unit cells until the difference in successive SGF values falls with in a certain tolerance limit calculated from the relative difference over the trace of $g_s$, typically set to $10^{-6}$. **n** iterations take into account **n** successive layers. Sancho-Rubio [10] proposed an algorithm where **n** iterations represent $2^n$ layers. As a result, this method provides a faster way of computing the SGF. Fig. 2 compares the two methods as a function of computation time of CNR width. With the implemented Sancho-Rubio method (0.25/0.034)~7 times larger CNR widths can be considered over typical methods. As a result, we could simulate for a large ribbon width (~200nm) to see quantum hall plateaus in low B-fields as we show in the following section. Typical end-to-end calculations for $R_H$ as a function of B-field with 1000 B-field points and a Graphene width of 100nm require ~70hrs on 20 state-of-the-art CPUs in an Intel EM64T machine [11].

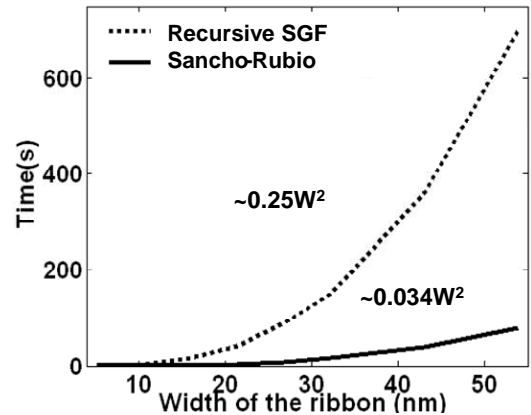

*Figure 2. Comparison of computation time required to calculate the self-energy of a contact in two different algorithms for increasing GNR width. Solid lines: surface Green function calculated with Sancho-Rubio iterative scheme. Dotted lines: surface Green function calculated with the conventional recursive algorithm.*

## 3. Results

The Quantum Hall effect is observed in a 2D sample at high magnetic (B) fields when Landau levels (LL) are formed. A LL is created at sufficient B-fields where the cyclotron orbit can form within the dimension. The larger the width, the smaller the required B-field to create the first LL. In Graphene, Landau energies are given by, $E_{LL}^n = \pm\sqrt{2e\hbar v_f^2 nB}$, where $v_f$ is the electron velocity in Graphene. Here we consider $v_f = 10^6$ m/s usually used in the literatures [2]. The 1st and 2nd LLs are at, 0.15eV and 0.23eV for B = 20T. Also there is a LL at E = 0, which is the zeroth LL [4, 5]. From the transmission characteristics in Fig. 3, we notice that both the transmission steps and DOS peaks are occurring at these LLs. However, we note that a high DOS present at E = 0 for ZGNR even without B-field. These are termed edge states [12] as they are spatially located at two edges of the sample. Both the semiconducting and metallic AGNRs show similar transmission properties under high B-field and we show only metallic case here.

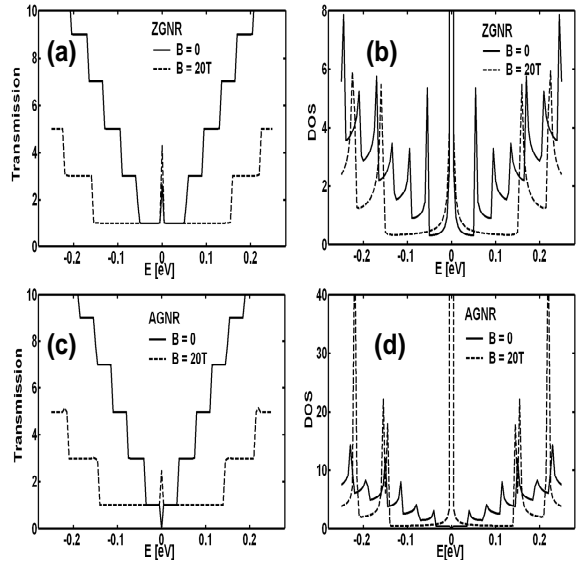

*Figure 3. Transmission and density of states of both ZGNR (a), (b), and AGNR (c), (d) in absence of magnetic (B) field (solid lines), in presence of B-field (dashed lines). Parameters: B = 20T, W = 53nm.*

The number of modes or the transmission steps in each LL is increasing in a sequence of odd integer numbers (2n+1), where n is an integer [Fig. 3 (a), (c)]. This is important because in the quantum Hall regime, the channel current is dependent on the number of current carrying modes when the Fermi level is lying between two LLs. In this case the current runs through the edges of the sample. At these energies +k states carry current through one side of the sample with an energy equal to the electrochemical potential of the left contact and –k states are running through the opposite edges of the sample with an energy equal to the electrochemical potential at the right contact. As a result, the channel becomes ballistic without scattering between the counter propagating edge states.

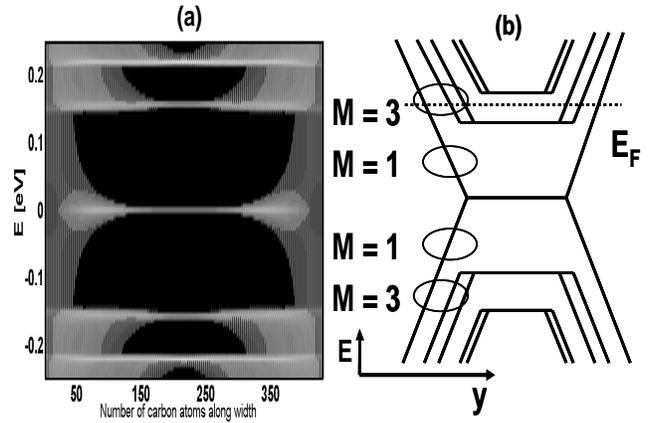

*Figure. 4. (a) Local DOS with energy, shows that the zeroth LL rising up (down) in energy for E>0 (E<0) near the edges providing an extra edge state at each edge (b) a schematic to illustrate the no. of edge states M at each energy. Parameters:53nm wide AGNR, B = 20T.*

Fig. 4 shows a local density of state plot in gray scale to demonstrate LL behavior near the edges of the sample where the potential becomes infinite. Fig. 4(a) shows the DOS at each LL along the width direction (y) at a particular B-field. The LL can be identified from the gray spots where there are high DOS. The zeroth LL increases (decreases) in energy near the edges due to hardwall potential at the width boundary. This level contributes an extra edge mode along with the two other edge modes from two inequivalent valleys at each LL. The schematic of Fig 4(a) shown in Fig. 4(b) clarifies the fact that when $E_F$ is in between two LLs the number of conducting edge states at each of the edge is increasing at steps of 1,3,5…(2n+1).

In Fig. 5(a) we show $R_H$ with increasing B-field for a ZGNR of 200nm ribbon width. $E_F$ has been fixed at 0.1eV. We notice that the plateaus are following a sequence, $R_H = h/(2e^2 M)$ where, $M = (2n+1)$. An increase in B-field increases the separation of LLs and hence the number of LLs below $E_F$ decrease and $R_H$ increases. At a particular B-field the number of edge states at $E_F$ defines the value of $R_H$.

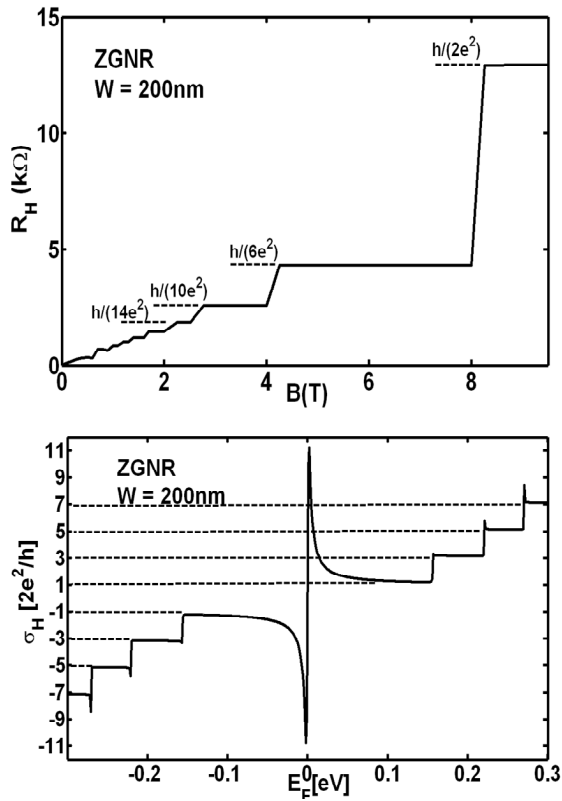

*Figure. 5. (a) Hall resistance ($R_H$) with different magnetic fields. 5(b) Hall conductivity ($\sigma_{xy}$) at varying $E_F$. Parameters: (a) 200nm wide ZGNR at $E_F = 0.1eV$, (b) 200nm wide ZGNR at $B = 20T$.*

Fig. 5(b) shows $\sigma_H$ at a particular B-field with varying $E_F$. At higher $E_F$ there is more edge modes and hence $\sigma_H$ increases. Here $\sigma_H$ is the conductivity per valley. As a result the plateaus are occurring at odd integer fillings. Here again, an odd number of edge modes at $E_F$ creating an odd integer QHE.

## 4. Conclusion

We have observed quantum Hall plateaus in the Graphene nano-ribbon system within the NEGF based quantum transport model in the ballistic regime. An efficient Sancho-Rubio algorithm has been used to calculate surface Green function for calculating contact self-energies. The magnetic field has also been incorporated in the contacts while calculating their self-energy matrices. This particular technique has relaxed the requirement of a large channel length to avoid abrupt transition in the magnetic potential profile from the contacts to the channel. As the contact self-energies are calculated with a fast iterative technique, we could simulate Graphene nano-ribbons of large width comparable to ones used in the experiments. Our numerical result shows exact qualitative agreement with the observed quantum Hall plateaus in the experiment.

**Acknowledgement:** nanoHUB computational resources were used in this work.

## References

1. Y. Zhang, *et al.*, "Experimental observation of the quantum Hall effect and berry's phase in Graphene", Nature **438**, 201 (2005).
2. C. Berger *etal*. "Electronic confinement and coherence in patterned epitaxial Graphene", Science **312**, 1191 (2006).
3. A. K. Geim and K. S. Novoselov, "The rise of Graphene", Nature mat. **6,** 181 (2007).
4. L. Brey and H. A Fertig, "Edge states and quantized Hall Effect in Graphene", Phys. Rev. B 73, 195408 (2006).
5. D. A. Abanin, P. A. Lee, L. S. Levitov, "Spin-filtered edge states and quantum-Hall effect in Graphene", Phys. Rev. Lett. **96**, 176803 (2006).
6. S. Datta, "Quantum Transport: Atom to Transistor", Cambridge University Press, Cambridge, (1997).
7. R. G- Mojarad and S. Datta, "Non-equilibrium Green's function based models for dephasing in quantum transport", Phys. Rev. B. **75,** 081301(R) (2007).
8. R Roger Lake, Gerhard Klimeck, R. Chris Bowen and Dejan Jovanovic, J. of Appl. Phys. **81**, 7845 (1997).
9. M. J. McLennan, Y. Lee, S. Datta, "Voltage drop in mesoscopic systems: A numerical study using quantum kinetic equation," Phys. Rev. B. **43,** 17 (1991).
10. M. P. L. Sancho, J. M. L. Sancho and J. Rubio J. Phys. F: Met. Phys. **14**, 1205 (1984).
11. nanoHUB.org computational resources.
12. K. Nakada *etal*. "Edge state in graphene ribbons: nanometer size effect and edge shape dependence", Phys. Rev. B. **54**, 17954 (1996).